# Vibron transport in macromolecular chains

Dalibor Čevizović[1,a], Zoran Ivić[1,b], Slobodanka Galović[1,c], Alexei Chizhov[2d] and Alexander Reshetnyak[3,e]

[1] University of Belgrade, Institute of Nuclear Sciences "Vinča" P.O. Box 522, Serbia

[2] Joint institute for Nuclear Research, Bogoliubov Laboratory of Theoretical Physics, Dubna, Russia

[3] Institute of Strength Physics and Materials Science of Siberian Branch Russian Academy of Sciences, Laboratory of Computer-Aided Design of Materials, Tomsk, 634021, Russia

[a]chevizd@vinca.rs, [b]zivic@vinca.rs, [c]bobagal@vinca.rs, [d]chizhov@theor.jinr.ru, [e]reshet@ispms.tsc.ru

**Keywords:** vibron excitation, phonon, vibron- phonon interaction, non adiabatic polaron, macromolecular chains.

**Abstract.** We study the hopping mechanism of the vibron excitation transport in the simple 1D model of biological macromolecular chains. We supposed that the vibron interaction with thermal oscillations of the macromolecular structural elements will result in vibron self -trapping, and the formation of the partial dressed vibron state. With use of the modified Holstein polaron model, we calculate vibron diffusivity in dependence of the basic system parameters and temperature. We obtain that the vibron diffusivity smoothly decreases in non adiabatic limit when the strength of the vibron-phonon coupling grows. However this dependence becomes by discontinuous one in case of growth of the adiabaticity of the system. The value of the critical point depends of the system temperature, and at room temperatures it belongs to the low or intermediate coupling regime. We discuss an application of these results to study of vibron transport to 3D bundles of such macromolecules chains considering it as polymer nanorods and to 2D polymer films organized from such macromolecules.

**Introduction**

There exist rigid arguments that the hydrolysis of adenosine triphosphate is the universal mechanism providing energy for diverse biological processes such as photochemical reactions, cross –membrane ion transfer and signal transduction, muscle contraction, cellular mobility, and transport, to mention just a few [1]. All these processes involve protein molecules as mediators of the long distance energy transfer. According to recent research [2,3,4], vibron self-trapping (ST) in protein macromolecules, and other biological macromolecular chains (MCs), will result in creation of partially dressed rather than fully dressed small polaron state (for solitonic nature of energy ST when the potential well may induced by local distortion of the molecular crystals, see e.g. [5]). As a consequence, the properties of ST vibron state may be quite different compared with the properties of the bare vibron excitation or fully dressed small polaron state. In fact, it was found that for the partially dressed vibrons there exist such values of the system parameters (critical values) for which the quasiparticle state abruptly changes from weakly dressed to heavy dressed, practically localized vibron state [2,3,6,7]. It was also found that the critical values of the system parameters are temperature dependent, and for some biological structures may belong to room temperature interval. This fact may significantly affect to the mechanism of the vibron motion through the MC.

In the paper, we investigate the temperature dependence of the diffusivity of the dressed vibron excitation states in MC for various values of the basic system parameters. We supposed that quasi-particle motion takes place via a sequence of random site–jumps, in each of which the quasiparticle hops to a first neighbor site of the MC. We also supposed that due to the vibron-phonon interaction, vibron ST process will results in formation of the partially dressed vibron states.

## Model Hamiltonian and Polaron Picture in Macromolecular Chain

As a theoretical basis, we consider Holstein molecular crystal Hamiltonian and take into account the vibron interaction with only optical phonon modes [8]. Latter assumption is justified because of the optical phonons play a crucial role in a process of the vibron dressing in hydrogen bonded MCs [9].

$$H = \Delta \sum_n A_n^+ A_n - J \sum_n A_n^+ (A_{n+1} + A_{n-1}) + \sum_q \hbar \omega_q B_q^+ B_q + \frac{1}{\sqrt{N}} \sum_n \sum_q F_q e^{iqnR_0} A_n^+ A_n (B_q + B_{-q}^+), \quad (1)$$

where $\Delta$, $\hbar$, $R_0$, N are respectively the vibron excitation energy, Planck constant, distance between two neighboring structural elements along the axis of MC and a number of structural elements, $A_n^+$ ($A_n$) describes the presence (absence) of the vibron quanta on the $n$-th structure element of the macromolecular chain, $B_q^+$ ($B_q$) creates (annihilates) phonon quanta with the frequency $\omega_q$, $J$ (~ -12 cm$^{-1}$) is inter–site overlap integral, which characterizes the vibron transfer between neighboring structure elements in the chain, and $F_q = F_{-q}^*$ is the vibron–phonon coupling parameter which governs the character of vibron ST states.

In order to find the most favorable vibron state, we follow the procedure which is described in [3]. Shortly, we reached partially dressed polaron picture with use of of the modified Lang–Firsov unitary transformation (MLFUT) for the Hamiltonian (1). Partial dressing approach can cover the entire region of the parameter space: from non adiabatic to adiabatic limits. It also can include the intermediate and weak coupling regimes [10]. From technical viewpoint, the method based on using of the MLFUT is variational method where the variational parameters determine the participation of the each phonon mode to the quasiparticle dressing. Doing so, we consider the variational method with single variational parameter, $0 \leq \delta \leq 1$ [3, 10], which is especially useful to derive the results due to its simplicity. This parameter may be used as a measure of the degree of the vibron dressing. In order to account for the influence of the thermal fluctuations on the properties of the ST vibron, we apply a simple mean field approach, based on the averaging of the transformed Hamiltonian over the phonon subsystem. An optimal state of the entire system in this case corresponds to a minimum of free energy for the system of MC, which we estimated by means of the Bogoliubov variational theorem [11]. According to the Bogoliubov theorem, the upper bound of the free energy for the system:

$$F_B = -k_B T \ln \sum_k e^{-E_{SP}/k_B T} + k_B T \sum_q \ln(1 - e^{-\hbar \omega_q / k_B T}), \quad (2)$$

where $E_{SP}(k) = \Delta - 1/N \sum_q \{(f_q - f_{-q}^*)F_q - \hbar \omega_q |f_q|^2\} - 2Je^{-W(T)} \cos(kR_0)$ is the variational energy for the small polaron, and $W(T) = 1/N \sum_q |f_q|^2 (2\bar{v} + 1)(1 - \cos(qR_0))$ is its vibron band narrowing factor, $\bar{v} = 1/(e^{\hbar \omega_q / k_B T} - 1)$ denotes the phonon average number at temperature $T$, and $f_q = \delta \frac{F_q^*}{\hbar \omega_q}$ is the parameter (depending on variational parameter δ), which characterizes the degree in which the vibron distorts the lattice and the lattice feedback on the vibron. In order to investigate the process of the vibron exciton transfer in MCs, we modify the model, developed by Holstein [8]. We supposed that vibron exciton can move along the MC by successive jumps from one to its first neighbor site in the chain. These jumps are accompanied by the emission or absorption of a number of phonons. In that case, the probability of the vibron jump to a neighboring MC site is:

$$P_{n \to n \pm 1}(T) = \frac{J^2}{\hbar^2} \sqrt{2\pi / I_1} \, e^{-I_2}, \quad (3)$$

where integrals $I_1$ and $I_2$ have the form:

$$I_1 = R_0 / \hbar^2 \pi \int_0^{\pi/R_0} 2 |F_q|^2 (1 - \cos qR_0) \operatorname{csch}(\hbar \omega_q / 2k_B T) dq \quad \text{and} \quad (4)$$

$$I_2 = R_0/\hbar^2\pi \int_0^{\pi/R_0} 2|F_q/\omega_q|^2 (1-\cos qR_0)\tanh(\hbar\omega_q/4k_BT)dq. \tag{5}$$

The further calculations should be made in terms of two independent dimensionless parameters: adiabatic parameter $B$, determining the character of the lattice deformation engaged in the polaron formation and coupling constant $S$ [2,3]. In the case when quasiparticle interacts with non dispersional optical phonon modes, they become: $B = 2J/\hbar\omega_0$, and $S = E_b/\hbar\omega_0$ (where $E_b = F^2/\hbar\omega_0$ is the lattice deformation energy, and $F = F_q = F^*_{-q}$). Finally, vibron relative diffusivity is determined as $D = \dfrac{P_{n\to n\pm 1}}{\omega_0}$ [8] in dimensionless form.

## Summary

The results obtained with use of the partial dressing method are presented on the Fig.1. and Fig.2. From Fig.1. one can see that a dependence of the variational parameter $\delta$ (i.e. vibron dressing) on system parameters in non adiabatic limit indicates that here we can get strongly dressed vibron – *polaron quasiparticle*. Only in the weak coupling or for intermediate coupling regimes, one can get *slightly lower quasiparticle dressing*. But with the growth of the system temperature, quasiparticle dressing increases. With increasing of $B$, and for low or intermediate values of $S$ the degree of the vibron dressing rapidly decreases (here we have *slightly dressed vibron states*). The dependence of the variational parameter $\delta$ is continuous function of $S$ for $B$ which belong to under –critical area of adiabatic parameter. But when $B$ reaches the critical value, $\delta$ abruptly changes at the point $S_C$.

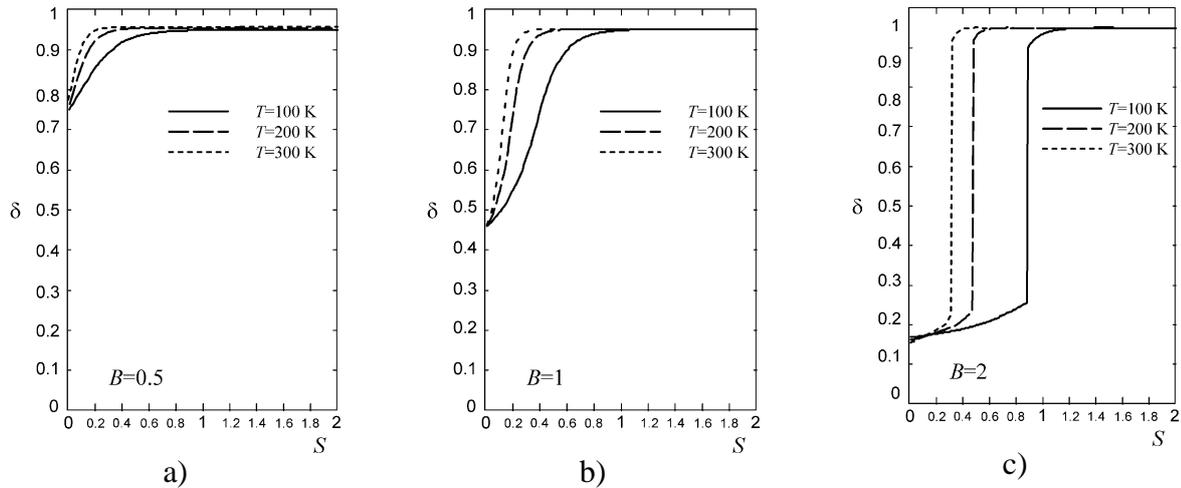

a)      b)      c)

Fig.1. The dependence of the variational parameter (i.e. vibron dressing) on the vibron-phonon coupling strength, for various temperatures and adiabatic parameter.

For the same values of the system parameters, relative diffusivity (i.e. the probability of the vibron hopping) significantly decreases with the growth of $S$ (Fig.2.). The shape of the $D(S)$ curve is smooth for under –critical values of $B$, and when $B = B_C$ it suffers interruption (relative diffusivity abruptly decreases). The decreasing trend of $D(S)$ curve can be understood as a consequence of the increasing of vibron dressing when $S$ grows. In the same time, diffusivity increases with the increasing of $B$. This behavior agrees with commonly adopted picture that in non adiabatic limit lattice deformation can capture quasiparticle and here we have strongly dressed, practically immobile quasiparticle, localized to one lattice site. Finally, it is interesting to remark that the vibron diffusivity decreases with the growth of the temperature for fixed values of $(S,B)$, when $S$ belongs to low or intermediate coupling regimes. At the first sight, this result is in contradiction with one obtained by Holstein [8], where non adiabatic polaron diffusivity increases with temperature. But, results in [8] are valid in non adiabatic and strong coupling regimes only.

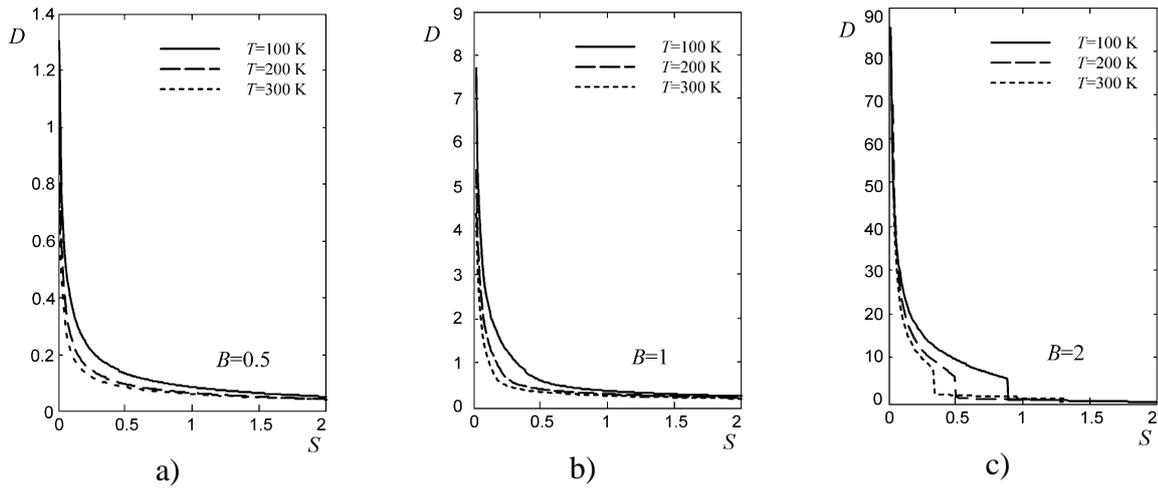

Fig.2. The dependence of the vibron hopping diffusivity on the vibron-phonon coupling strength, for various temperatures and adiabatic parameter.

Due to the fact that the difference among the values of the transfer integral which determines the vibron hopping along the single MC, and the transfer integral which determines the vibron hopping between two chains in complex structure, it is possible to occur that the vibron exciton can have high mobility only in one direction, while it is practically immobile in other directions.

It is very interesting to extend the results of this study of vibron localization and transport from single protein-like MC both to 3D bundle as polymer nanorod and to 2D polymer film of such MC. The work was supported by the Serbian Ministry of Education and Science under Grant No III-45010, and the bilateral project between Serbian Ministry of Education and Science and JINR Dubna "Theory of Condensed Matter"

**References**

[1]  A S. Davydov, Biologya i Kvantovaya Mehanika, Naukova Dumka, Kiev (in Russian) 1979.

[2]  D. Čevizović, S. Galović, A. Reshetnyak, Z. Ivić, The vibron dressing in $\alpha$-helicoidal macromolecular chains, Chin. Phys. B, 22 (2013) 060501.

[3]  D. Čevizović, S. Galović, Z. Ivić, Nature of the vibron self –trapped states in hydrogen – bonded macromolecular chains, Phys. Rev. E 84 (2011) 011920.

[4]  V. Pouthier, Vibron phonon in a lattice of H-bonded peptide units: A criterion to discriminate between the weak and the strong coupling limit, Journ. Chem. Phys., 132 (2010) 035106.

[5]  A S. Davydov, Solitons in Molecular Systems, Physica Scripta, 20 (1979) 387-394.

[6]  G. Iadonisi, V. Cataudella, G. de Filipis, and D. Ninno, Coexistence of large and small mass polarons, Europhys. Lett., 41 (1998) 309 -314.

[7]  P. Hamm and G. P. Tsironis, Barrier crossing to the small Holstein Polaron regime, Phys. Rev. B, 78 (2008) 092301.

[8]  T. Holstein, Studies of Polaron Motion, Annals of Physics, 281 (2000) 725-773.

[9]  D. M. Alexander and J. A. Krumhansl, Localized excitations in hydrogen bonded molecular chains, Phys. Rev. B, 33 (1986) 7172-7185.

[10] D. W. Brown and Z. Ivić, Unification of Polaron and Soliton Theories of Exciton Transport, Phys. Rev. B, 40 (1989) 9876-9887.

[11] I.A. Kvasnikov, Kvantovaya statistika, URSS, Moskva, 2011. (in Russian).